\documentclass[12pt,a4paper]{article}
\usepackage{epsfig}
\newcommand{\am}{\alpha_{m\! a\! x}}
\newcommand{\bra}[2]{\left\langle{#1}|{#2}\right\rangle}
\newcommand{\Coo}[2]{W^{{#1},{#2}}}

\newcommand{\Cou}[2]{V^{{#1},{#2}}}
\newcommand{\cou}[3]{V_{{#1},{#2}}^{({#3})}}
\newcommand{\dil}{\left({\mathcal D}\phi,\frac{\delta}{\delta\phi}
   \right)}
\newcommand{\eig}[1]{W^{({#1})}(\phi)}
\newcommand{\fix}[1]{V^{({#1})}(\phi)}
\newcommand{\Ind}{{\mathcal I}}
\newcommand{\Ker}{{\mathcal P}}
\newcommand{\la}[1]{\lambda^{#1}}
\newcommand{\lap}[1]{\left(\frac{\delta}{\delta\phi_1},{#1}
   \frac{\delta}{\delta\phi_2}\right)}
\newcommand{\mc}{\multicolumn}
\newcommand{\NNN}{{\bf N}}
\newcommand{\obs}[2]{{\mathcal O}_{{#1},{#2}}(\phi)}
\newcommand{\Obs}[1]{{\mathcal O}_{{#1}}(\phi)}
\newcommand{\pade}{Pad\'{e} }
\newcommand{\pot}{V(\phi)}
\newcommand{\Pro}[1]{{\mathcal P}_{{#1}}}
\newcommand{\pro}[2]{{\mathcal P}_{{#1},{#2}}}
\newcommand{\Ref}[1]{(\ref{#1})}
\newcommand{\spe}[1]{\lambda^{({#1})}}
\newcommand{\ZZZ}{{\bf Z}}
\begin{document}
\begin{titlepage}
\begin{center}
{\Large  
Interpolation Parameter and Expansion for the Three Dimensional 
Non--Trivial Scalar Infrared Fixed Point
} \\[10mm]
\end{center}
\begin{center}
{\Large 
C. Wieczerkowski\,$^1$ 
and J. Rolf\,$^2$ 
} \\[10mm]
\end{center}
\begin{center}
$^1$\,Institut f\"ur Theoretische Physik I,
Universit\"at M\"unster,\\
Wilhelm-Klemm-Stra\ss e 9, D-48149 M\"unster,\\
wieczer@yukawa.uni-muenster.de \\[2mm]
$^2$\,Niels Bohr Institute, \\
Blegdamsvej 17, DK-2100 Copenhagen,\\
rolf@alf.nbi.dk \\[2mm]
\end{center}
\vspace{-10cm}
\hfill\parbox{35mm}{MS-TPI 96-10\\NBI-HE-96-34}
\vspace{11cm}
\begin{abstract}
We compute the non--trivial infrared $\phi^4_3$--fixed point by means
of an interpolation expansion in fixed dimension. The expansion is 
formulated for an infinitesimal momentum space renormalization group.
We choose a coordinate representation for the fixed point interaction
in derivative expansion, and compute its coordinates to high orders by 
means of computer algebra. We compute the series for the critical 
exponent $\nu$ up to order twenty five of interpolation expansion 
in this representation, and evaluate it using \pade, Borel--\pade, 
Borel--conformal--\pade, and Dlog--\pade resummation. The resummation 
returns  $0.6262(13)$ as the value of $\nu$.
\end{abstract}
\end{titlepage}
\section{Introduction}

Non--trivial fixed points are a highly challenging aspect of renormalization
theory. Much of what is known about non--trivial fixed point is due to the
$\epsilon$--expansion of Wilson and Fisher \cite{WF72,WK74}, which is an 
interpolation from a critical dimension to the one of interest. In this 
paper we present another interpolation scheme where the dimension of the 
underlying (Euclidean) space--time is kept fixed. 

A prototype of a non--trivial fixed point is the infrared fixed point of 
massless $\phi^4$--theory in three dimensions, also called Wilson fixed
point \cite{WK74}. We choose it as an example for our method. Although it 
has been investigated by various other means, for instance by Monte Carlo 
simulations of the three dimensional Ising model close to criticality, 
hopping parameter expansion, field theoretic perturbation theory for its 
scaling limit, and numerical integration of renormalization group flows 
in a number of setups, our knowledge of it is far from satisfactory.   
Accurate data for its spectrum of anomalous dimensions is lacking, its 
functional form is largely unknown, in particular its locality properties, 
and its mathematical construction remains an outstanding difficult problem. 
We mention \cite{ZJ89} and references therein as a guide to the extensive 
literature. We mention further that all this has been accomplished to a 
very satisfactory status in the hierarchical approximation by Koch and 
Wittwer \cite{KW91}. Our interpolation is a brick in the analysis of the 
full model. 

As starting point we choose a functional differential equation from the
infinitesimal renormalization group of Wilson \cite{WK74}. Specifically 
we choose a normal ordered and rescaled representation for the fixed point 
interaction, expressed in terms of a scalar field with non--anomalous 
scaling dimension. It contains a bilinear renormalization form. This 
bilinear form is continuously turned on with an auxiliary parameter such
that zero gives a linear theory and one restores the full equation. The
linear theory is arranged such that the $\phi^4$--interaction acquires
the scaling dimension zero in it. We then expand the fixed point interaction
into a power series in the interpolation parameter. This part is similar
to the $\epsilon$--expansion. In order to perform the expansion to high 
orders on the computer, the interaction is written in a basis of 
interactions which includes a general two point interaction in
derivative expansion together with local higher interactions. We compute
both the fixed point interaction and the eigenvalue associated with a
massive perturbation. The resulting power series are evaluated by means of
\pade, Dlog, Borel--\pade, and Borel--conformal--\pade resummation. 

The interpolation idea applies also to other renormalization schemes.
The infinitesimal renormalization group is a particularly  convenient one 
because it involves a minimal set of Feynman integrals. Interpolations
in fixed dimensions can also be formulated for discrete renormalization 
group transformations both in continuum regularization and on the lattice, 
at the expense of dealing with general non--linear rather than quadratic 
equations. It is conceivable that our interpolation can be given a meaning 
beyond perturbation theory. 

The paper is organized as follows. In section two we explain the structure 
of our particular functional differential equation. It is taken from 
\cite{W96} and is the Wilson equation \cite{WK74} in a kind
of interaction picture. In section three we present our interpolation 
scheme. It is compared to a naive interpolation which has only a 
trivial solution. We solve our equations to lowest order to explain 
their recursive treatment. In section four we discuss their form in 
a coordinate representation. The result is a set of algebraic recursion 
relations for the fixed point interaction. They involve a set of structure 
constants whose computation again involves certain Feynman integrals 
and multiplicities. We devote section five to this issue. 
In section six and seven the eigenvalue problem for the scaling fields
of the non--trivial fixed point and their anomalous dimensions is 
treated along the same lines. We restrict our attention to a mass
perturbation of the fixed point and the associated critical index $\nu$. 
In chapter eight and nine the resulting recursions are studied 
by means of computer algebra. We conclude with a brief discussion of
our results on the value $\nu$.
\section{Renormalization group fixed point}
We consider a real scalar field $\phi$ on three dimensional Euclidean
space. We use a momentum space renormalization group built from the
decomposition of a massless propagator $v$ with exponential ultraviolet
regulator. The renormalization group will be formulated in terms of an
interaction $V(\phi)$. Concerning the general background, we refer to
the work of Wilson \cite{W71}, Wilson and Kogut \cite{WK74}, and also
to Gallavotti \cite{G85}. Our setup will be identical with that in
\cite{W96}.

We study the non--trivial infrared fixed point in three dimensions as
solution to the functional differential equation
\begin{equation}
\dil\pot=\bra{\pot}{\pot},
\label{fixed}
\end{equation}
which was derived in \cite{W96}.
Eq. \Ref{fixed} is a normal ordered and rescaled variant of the
infinitesimal renormalization group due to Wilson \cite{WK74}.
Its origin is a flow equation governing the behaviour of interactions 
upon the infinitesimal change of a floating cutoff. Eq. \Ref{fixed} 
gives stationary flows modulo the rescaling of units. 
The left hand side of 
\Ref{fixed} is a generator of dilatations
\begin{equation}
\dil\pot=\frac{{\rm d}}{{\rm d}L}V(\phi_L)
\biggr\vert_{L=1},\quad \phi_L(x)=L^{-1/2}\phi\left(\frac{x}{L}
\right),
\end{equation}
acting on the interaction. The field $\phi$ is here rescaled
non--anomalously with its canonical dimension at the trivial fixed
point. It should be distinguished from the scaling fields of the
infrared fixed point which have non--zero anomalous dimensions.
The right hand side of \Ref{fixed} is a bilinear renormalization
group form
\begin{equation}
\bra{\pot}{\pot}=\lap{\chi}\exp\left\{
\lap{v}\right\}V(\phi_1)V(\phi_2)\biggr\vert_{\phi_1=\phi_2=\phi}.
\label{bilinear}
\end{equation}
It can be visualized as a sum of contractions between two copies
of the interaction. Each contraction is made of one {\sl hard}
propagator $\chi$ and any number of {\sl soft} propagators $v$.
The propagators are here given by
\begin{equation}
\widetilde{\chi}(p)=e^{-p^2},\quad
\widetilde{v}(p)=\frac{e^{-p^2}}{p^2},
\label{propagators}
\end{equation}
as in \cite{W96}. Eq. \Ref{fixed} is the main dynamical equation
in this investigation. 
Being a differential equation, it has to be
supplied with further data to select a particular solution. In a
rigorous theory in the sense of Glimm and Jaffe \cite{GJ87}, the
infrared fixed point should come as a global $\ZZZ_2$--symmetric
solution, where {\sl global} refers to some criterion of finiteness.
Our point of view in this approach will be more modest. An
interaction $V(\phi)$ will stand for a power series
\begin{equation}
V(\phi)=\sum_{n=1}^{\infty}\int{\rm d}^3x_1
\cdots{\rm d}^3x_{2n}\;\phi(x_1)\cdots\phi(x_{2n})\;
V_{2n}(x_1,\ldots,x_{2n})
\label{power}
\end{equation}
in the field, with symmetric Euclidean invariant distributional
kernels given by Fourier integrals
\begin{eqnarray}
V_{2n}(x_1,\ldots,x_{2n})&=&\int
\frac{{\rm d}^3p_1}{(2\pi)^3}\cdots
\frac{{\rm d}^3p_{2n}}{(2\pi)^3}\;e^{i(p_1x_1+\cdots+p_{2n}x_{2n})}
\nonumber\\&&(2\pi)^3\delta^{(3)}(p_1+\cdots+p_{2n})
\widetilde{V}_{2n}(p_1,\ldots,p_{2n})
\end{eqnarray}
of smooth momentum space kernels. I.e., we identify an interaction
with its collection of momentum space kernels. The question of
convergence of the expansion \Ref{power} in powers of fields will
not be addressed. It is conceivable that it could be tackled
with a suitable norm on the collection of momentum space kernels
as a whole. 

In the iterative approach to be defined below we will
meet at finite order no more than polynomial expressions in the
field. We will understand \Ref{fixed} as a system of differential
equations for the momentum space kernels. Its explicit form
can be looked up in \cite{W96}. Boundary data is substituted for
by the condition of regularity. Homogeneous functions give particular
kernels, which correspond to scaling fields of the trivial fixed
point. Expanding a kernel in powers of momentum derivatives, we can
always express it in terms of such scaling fields. To distinguish
them from the scaling fields of the non--trivial fixed point and
also because we will use perturbation theory, we will speak of them
as {\sl vertices}.
\section{Interpolation parameter and expansion}
Our strategy to solve \Ref{fixed} is to interpolate to a solvable
situation. If the interpolation is smooth it can be performed by
means of perturbation theory. A natural candidate
is
\begin{equation}
\dil V(\phi,z)=z\bra{V(\phi,z)}{V(\phi,z)}
\label{naive}
\end{equation}
with an interpolation parameter $z=0\ldots 1$. It can be thought
to turn on continuously the bilinear form, which is identified as
the source of troubles. The interpolation \Ref{naive} is inappropriate
for the following reason, when the dimension parameter is fixed
to three. Expand the interpolated interaction as a function of the
interpolation parameter in a power series
\begin{equation}
V(\phi,z)=\sum_{r=0}^\infty z^r\;V^{(r)}(\phi).
\label{pertur}
\end{equation}
Unfortunately there is little hope that \Ref{pertur} has a finite
radius of convergence both in the case of \Ref{naive} and the
interpolation \Ref{better} considered below. To be cautious we will
therefore view \Ref{pertur} as a formal power series and interpret
all equations below in this sense. It will however be argued that
non--perturbative information can be extracted by Borel
resummation. In order to solve \Ref{naive}, the expansion
\Ref{pertur} has to satisfy
\begin{equation}
\dil\fix{r}=\sum_{s=0}^{r-1}\bra{\fix{s}}{\fix{r-1-s}}
\label{perfix}
\end{equation}
to every order $r\in\NNN$, with the understanding $\fix{-1}=0$.
In particular it requires the interaction to satisfy
\begin{equation}
\dil\fix{0}=0
\label{marginal}
\end{equation}
to zeroth order. In other words, the zeroth order has to be
a marginal scaling field of the trivial fixed point. In three
dimensions we have two marginal scaling fields, a wave
function term and a $\phi^6$--vertex, to be abbreviated as
\begin{equation}
{\mathcal O}_{1,1}(\phi)=\int{\rm d}^3x\;\phi(x)
(-\bigtriangleup )\phi(x),
\quad {\mathcal O}_{3,0}(\phi)=\int{\rm d}^3x
\;\phi(x)^6.
\label{vertices}
\end{equation}
We emphasize that vertices should be understood as momentum space
kernels at zero momentum and their Taylor expansions. Each of
them comes with a formal orthogonal projector $\pro{1}{1}$ and
$\pro{3}{0}$, selecting the corresponding vertex from a general
interaction \Ref{power}. The zeroth order has to be a linear
combination
\begin{equation}
\fix{0}=V^{(0)}_{1,1}\;{\mathcal O}_{1,1}(\phi)+
V^{(0)}_{3,0}\;{\mathcal O}_{3,0}(\phi).
\label{zero}
\end{equation}
The coupling constants are not determined by the zeroth order
equation \Ref{marginal}. To first order, \Ref{perfix} reads
\begin{equation}
\dil\fix{1}=\bra{\fix{0}}{\fix{0}}.
\label{problem}
\end{equation}
But eqs. \Ref{zero} and \Ref{problem} together have only a
trivial solution. The left hand side of \Ref{problem} cannot
contain the vertices \Ref{vertices} because the dilatation generator
$\dil$ has no marginal image. Therefore, \Ref{problem} requires
that
\begin{equation}
\pro{1}{1}\bra{\fix{0}}{\fix{0}}=
\pro{3}{0}\bra{\fix{0}}{\fix{0}}=0.
\end{equation}
Computing the bilinear form with two copies of \Ref{zero}
inevitably gives $V^{(0)}_{1,1}=V^{(0)}_{3,0}=0$.
Eq. \Ref{naive} is thus an inappropriate interpolation and has
to be given up.

A way around the obstacle is to interpolate
simultaneously the dimensionality of the theory. This is the
strategy of the $\epsilon$--expansion of Wilson and Fisher
\cite{WF72} in a field theoretic setup. Another way is to
interpolate the scaling dimension, remaining firmly in three
dimensions. We choose this second route and replace \Ref{naive}
by
\begin{equation}
\left[\dil-1\right] V(\phi,z)=
z\;\bra{V(\phi,z)}{V(\phi,z)}-z\;V(\phi,z).
\label{better}
\end{equation}
The power series expansion \Ref{pertur} solves \Ref{better}
if the coefficients satisfy the system of differential equations
\begin{equation}
\left[\dil-1\right]\fix{r}=
\sum_{s=0}^{r-1}\bra{\fix{s}}{\fix{r-1-s}}-\fix{r-1}
\label{system}
\end{equation}
to all orders $r\in\NNN$. To order zero, \Ref{system} requires
now
\begin{equation}
\left[\dil-1\right]\fix{0}=1
\label{unit}
\end{equation}
in contrast to \Ref{marginal}. The zeroth order interaction is
now a scaling field with unit scaling dimension. In three dimensions
we have only one candidate, the $\phi^4$--vertex
\begin{equation}
{\mathcal O}_{2,0}(\phi)=\int{\rm d}^3x\;\phi(x)^4.
\label{phifour}
\end{equation}
The zeroth order interaction thus has to be proportional to
\Ref{phifour}. The proportionality factor is the $\phi^4$--coupling.
It is not determined by the zeroth order equation \Ref{unit}.
We conclude that
\begin{equation}
\fix{0}=\cou{2}{0}{0}\;{\mathcal O}_{2,0}(\phi).
\end{equation}
This expansion proves to have indeed a non--trivial solution. To
see this, consider the first order equation in \Ref{system}.
It reads
\begin{equation}
\left[\dil-1\right]\fix{1}=
\bra{\fix{0}}{\fix{0}}-\fix{0}.
\label{first}
\end{equation}
Eq. \Ref{first} cannot have a $\phi^4$--vertex on its left hand
side. Therefore it is required that
\begin{equation}
\pro{2}{0}\biggl\{\bra{\fix{0}}{\fix{0}}-\fix{0}\biggr\}=0.
\end{equation}
Computing the bilinear form, this condition reads explicitely
\begin{equation}
\cou{2}{0}{0}\biggl\{144\;\widetilde{\chi}\star\widetilde{v}(0)
\;\cou{2}{0}{0}-1\biggr\}=0,
\end{equation}
where $\star$ means convolution times $(2\pi)^{-3}$.
Besides the trivial solution $\cou{2}{0}{0}=0$ it has a
non--trivial solution
\begin{equation}
\cou{2}{0}{0}=
\frac{1}{144\widetilde{\chi}\star\widetilde{v}(0)}=
\frac{(2\pi)^{3/2}}{72}=0.21874445\ldots
\label{nontrivial}
\end{equation}
The value of the $\phi^4$--coupling at any given order will in fact
be determined by the equations at the next order, a feature of this 
particular interpolation expansion. To first order, the interaction 
can be split into
\begin{equation}
\fix{1}=\cou{2}{0}{1}\obs{2}{0}+
\pro{2}{0}^\perp\fix{1}
\end{equation}
with $\pro{2}{0}^\perp=1-\pro{2}{0}$ the projector
on the formal orthogonal complement. Eq. \Ref{first} defines a
system of first order differential equations for the momentum space
kernels therein. They have a unique integral in the space of smooth
functions of momenta, see \cite{W96}. We denote this integral by
\begin{equation}
\pro{2}{0}^{\perp}\fix{1}=
\left[\dil-1\right]^{-1}
\pro{2}{0}^{\perp}\bra{\fix{0}}{\fix{0}}.
\label{integralone}
\end{equation}
This iterative scheme carries on to every order of interpolation
expansion. Consider eq. \Ref{system} at order $r\geq 2$.
The first step of the iteration is to compute $\cou{2}{0}{r-1}$
at order $r-1$. Making use of \Ref{nontrivial}, its value follows 
from
\begin{eqnarray}
\cou{2}{0}{r-1}\;\obs{2}{0}&=&
-2\pro{2}{0}\bra{\fix{0}}{\pro{2}{0}^\perp\fix{r-1}}-
\nonumber\\& &\quad
\sum_{s=1}^{r-2}\pro{2}{0}\bra{\fix{s}}{\fix{r-1-s}}.
\label{coupiteration}
\end{eqnarray}
Again one splits the order $r$ interaction into
\begin{equation}
\fix{r}=\cou{2}{0}{r}\;\obs{2}{0}+
\pro{2}{0}^{\perp}\fix{r}
\end{equation}
and computes the formal orthogonal complement to the
$\phi^4$--vertex by integrating the first order differential
equations \Ref{system}. The result can be written as
\begin{eqnarray}
\pro{2}{0}^{\perp}\fix{r}&=&
\left[\dil-1\right]^{-1}
\Biggl\{\sum_{s=0}^{r-1}\pro{2}{0}^{\perp}
\bra{\fix{s}}{\fix{r-1-s}}-
\nonumber\\& &\quad
\pro{2}{0}^{\perp}\fix{r-1}\Biggr\}.
\label{otheriteration}
\end{eqnarray}
Thereafter it is time to proceed to the equations at order
$r+1$. For this scheme to work as above it is important
that the kernel of $\dil -1$ be one dimensional. Otherwise we
would have to compute further order $r-1$ data from the
equations to order $r$. An example where this happens is the
$\phi^4$--trajectory in four dimensions \cite{W96}.

Although being in principle doable, the computation of this
scheme to very high orders of interpolation expansion is a
tedious enterprise. The main work is the computation of a wealth
of Feynman kernels generated in the course of iteration.
A low order analysis of this program will be presented
elsewhere. In this paper we choose to evaluate the expansion
to high orders for a sub--class of contributions in the
iteration. For this purpose we reformulate the fixed point
equation \Ref{better} into an algebraic system of equations for
a set of coupling constants. We find it interesting by its own.
It also allows to perform the expansion on a computer.
\section{Coordinate representation}
We choose a system of vertices $\Obs{i}$ labelled by elements $i$
of an index set $\Ind$. The vertices will be required to be
$\ZZZ_2$--symmetric, Euclidean invariant, and linearly independent.
They will also be required to be regular in the sense that they are
given by smooth momentum space kernels. We choose the system such
that the dilatation generator acts linearly on it through a scaling
dimension matrix
\begin{equation}
\dil\Obs{i}=\sum_{j\in\Ind}\Obs{j}\;\Sigma^{j}_{i}.
\label{scalematrix}
\end{equation}
We restrict our attention to systems with the property that the
scaling dimension matrix is diagonalizable. Non--diagonalizable
matrices will not be considered here. In this case, we can
arrange the system to consist of eigenvectors
\begin{equation}
\dil\Obs{i}=\sigma_{i}\Obs{i}.
\label{scalefield}
\end{equation}
In other words, we take $\Obs{i}$ to be a scaling field of the
trivial fixed point with scaling dimension $\sigma_i$. Recall that
such vertices are given by homogeneous momentum space kernels.
Eq. \Ref{scalematrix} says that the system closes under the action
of an infinitesimal dilatation. We also require it to close under
the action of the bilinear renormalization group form. For any two
vertices $\Obs{i}$ and $\Obs{j}$ the bilinear form \Ref{bilinear}
will be assumed to be a linear combination
\begin{equation}
\bra{\Obs{i}}{\Obs{j}}=
\sum_{k\in\Ind}\Obs{k}F^{k}_{i,j}
\label{structure}
\end{equation}
with a set of structure constants $F^{k}_{i,j}$. The scaling
dimensions $\sigma_i$ and the structure constants $F^{k}_{i,j}$
comprise all the information needed in the following about the
system of vertices. We remark that the structure constants are
well defined through
\begin{equation}
F^{k}_{i,j}\;\Obs{k}=
\Pro{k}\bra{\Obs{i}}{\Obs{j}}
\end{equation}
even when the system does not close under the bilinear form.
In this case \Ref{structure} holds only up to an error term.
Below we will indeed work with an approximation of this kind
and argue that the error term is small.

We define a coordinate representation for the interpolated
interaction in terms of a given system of vertices as
\begin{equation}
V(\phi,z)=\sum_{i\in\Ind}\Obs{i}\; V^{i}(z).
\label{coordinates}
\end{equation}
The idea is then to investigate the interpolation
\Ref{better} for the infrared fixed point by means of the
parameter dependent coordinates \Ref{coordinates}. Eq.
\Ref{better} becomes a system of algebraic equations
\begin{equation}
(\sigma_{k}-1)\; V^{k}(z) =
z\sum_{i,j\in\Ind} F^{k}_{i,j}\;V^{i}(z)\;V^{j}(z)-
z\;V^{k}(z)
\label{algebra}
\end{equation}
in the coordinate representation. The advantage of \Ref{algebra}
as compared to \Ref{better} is that we are no longer dealing with
differential equations for momentum space kernels. Their
integration is hidden in the structure constants. If the
interpolation is smooth, we can expand the coordinate functions
into power series
\begin{equation}
V^{k}(z)=\sum_{r=0}^{\infty}\;z^r\;\Cou{k}{r}.
\label{coeffs}
\end{equation}
By standard arguments \Ref{coeffs} is expected to be singular but
Borel summable. Our below evaluation of \Ref{coeffs} supports this
expectation. Eq. \Ref{coeffs} yields a solution to \Ref{algebra} 
in the sense of a formal power series in $z$ if the coefficients
obey
\begin{equation}
(\sigma_{k}-1)\;\Cou{k}{r} =
\sum_{s=0}^{r-1}\sum_{i,j\in\Ind} F^{k}_{i,j}\;
\Cou{i}{s}\;\Cou{j}{r-1-s}-
\Cou{k}{r-1}
\label{computer}
\end{equation}
holds for all couplings $k\in\Ind$ to all orders $r\in\NNN$ of
interpolation expansion. We organize \Ref{computer} into a
recursion relation which can be solved on the computer.
To zeroth order \Ref{computer} simplifies to the linear
equation
\begin{equation}
(\sigma_{k}-1)\;\Cou{k}{0}=0.
\label{orderzero}
\end{equation}
We assume that our system of vertices contains only
one element labelled by $k=\underline{2}=(2,0)$ such that
$\sigma_{\underline{2}}=1$. This element is of course the
$\phi^4$--vertex \Ref{phifour}. All other elements are
assumed to have scaling dimensions different from one.
Then \Ref{orderzero} implies that
\begin{equation}
\Cou{k}{0}=\Cou{\underline{2}}{0}
\;\delta_{\underline{2},k}.
\end{equation}
The value of $\Cou{\underline{2}}{0}$ is as above determined
by \Ref{computer} to order one,
\begin{equation}
(\sigma_{k}-1)\;\Cou{k}{1}=
\Cou{\underline{2}}{0}
\left(F^{\underline{2}}_{\underline{2},\underline{2}}\;
\Cou{\underline{2}}{0}-\delta_{\underline{2},k}\right).
\label{orderone}
\end{equation}
Evaluating \Ref{orderone} for $k=\underline{2}$ it follows
immediately that we have
\begin{equation}
\Cou{\underline{2}}{0}=
\frac{1}{F^{\underline{2}}_{\underline{2},\underline{2}}},
\label{coupzero}
\end{equation}
besides the trivial solution $\Cou{\underline{2}}{0}=0$.
Eq. \Ref{orderone} does not tell the value of
$\Cou{\underline{2}}{1}$. But for $k\in\Ind\setminus
\{\underline{2}\}$ it gives
\begin{equation}
\Cou{k}{1}=
\frac{F^{k}_{\underline{2},\underline{2}}
\;\left(\Cou{\underline{2}}{0}\right)^2}
{\sigma_{k}-1}.
\label{otherone}
\end{equation}
Eq. \Ref{coupzero} and \Ref{otherone} are of course the
coordinate versions of \Ref{nontrivial} and \Ref{integralone}.
The strategy to any order $r>1$ is again to first compute
$\Cou{\underline{2}}{r-1}$ and thereafter
$\Cou{k}{r}$ for $k\in\Ind\setminus\{\underline{2}\}$.
The explicit formulas are
\begin{equation}
\Cou{\underline{2}}{r-1}=
-2\sum_{i\in\Ind\setminus\{\underline{2}\}}
F^{\underline{2}}_{i,\underline{2}}
\;\Cou{\underline{2}}{0}\;\Cou{i}{r-1}-
\sum_{s=1}^{r-2}\sum_{i,j\in\Ind}
F^{\underline{2}}_{i,j}\;
\Cou{i}{s}\;\Cou{j}{r-1-s}
\label{couprecursion}
\end{equation}
and
\begin{equation}
\Cou{k}{r}=
\frac{1}{\sigma_{k}-1}\left\{
\sum_{s=0}^{r-1}\sum_{i,j\in\Ind}
F^{k}_{i,j}\;\Cou{i}{s}\;\Cou{j}{r-1-s}-
\Cou{k}{r-1}\right\}
\label{otherrecursion}
\end{equation}
in complete analogy to \Ref{coupiteration} and \Ref{otheriteration}.
Thus once we know the scaling dimensions and the
structure constants, the iteration proceeds by means of purely
algebraic operations. We remark that the sums in \Ref{couprecursion}
and \Ref{otherrecursion} will be finite in the system of vertices
considered below. The reason is that the outcome of the
bilinear form of two monomials in the field is a polynomial in
the field of finite order, and consists only of connected
vertices. A very interesting question is whether it is possible
to find finite systems of vertices that close under both
\Ref{scalematrix} and \Ref{structure}. It is clear that this
cannot be achieved in terms of polynomial vertices. Unfortunately
no such system is known in three dimensions.
\section{Structure constants}
We consider the following system of vertices. First we include
a full two point vertex in derivative expansion. A convenient
notation for it is
\begin{equation}
\obs{1}{\alpha}=
\int{\rm d}^3x\;\phi(x)\;
(-\bigtriangleup)^{\alpha}\;\phi(x),
\label{twovertex}
\end{equation}
where $\alpha =0,1,2,\ldots$. Second we include local $(2n)$--point
vertices with arbitrary many external legs. They will be abbreviated
as
\begin{equation}
\obs{n}{0}=\int{\rm d}^3x\;\phi(x)^{2n},
\label{anyvertex}
\end{equation}
where $n=2,3,4,\ldots$. Notice that both \Ref{twovertex} and
\Ref{anyvertex} meet the demands stated at the beginning of the
previous section. More general interactions include also
momentum dependent higher vertices \Ref{anyvertex}. They will
not be considered here. Our index set is thus
\begin{equation}
\Ind =\{1\}\times\{\alpha\in\NNN\vert\alpha\geq 0\}\cup
\{n\in\NNN\vert n\geq 2\}\times\{0\}
\end{equation}
and $\underline{2}=(2,0)$. The bilinear form does {\sl not}
close under this set of vertices. For instance two local
vertices \Ref{anyvertex} contract in general to a bilocal
vertex. Thus if we perform an iteration \Ref{couprecursion} and
\Ref{otherrecursion} with this system of vertices, we make
a systematic error due to the truncation of the system. Our
ansatz rests upon the assumption that non--local higher vertices
are small compared to their local parts. 

The scaling dimensions of \Ref{twovertex} and \Ref{anyvertex} come 
out as
\begin{equation}
\sigma_{1,\alpha}=2-2\alpha ,\quad
\sigma_{n,0}=3-n.
\end{equation}
The structure constants for this set of vertices come out as follows.
Two quadratic vertices always contract again to a quadratic vertex. 
The associated structure constants are computed to
\begin{equation}
F^{(1,\alpha)}_{(1,\beta),(1,\gamma)}=
4\;\frac{(-1)^{\alpha-\beta-\gamma}}{(\alpha-\beta-\gamma)!}
\;\Theta_{\alpha,\beta+\gamma},
\label{structureone}
\end{equation}
where $\Theta_{a,b}=1$ for $a\geq b$ and zero else. A quadratic
vertex and a higher vertex return upon pairing both a quadratic
vertex and a higher vertex. First we have
\begin{equation}
F^{(1,\gamma)}_{(1,\alpha),(2,0)}=
24\;\widetilde{K}_{1,\alpha}(0)\;\delta_{\gamma,0}.
\label{structuretwo}
\end{equation}
The structure constant \Ref{structuretwo} involves the one
loop integral
\begin{equation}
\widetilde{K}_{1,\alpha}(p)=
\int\frac{{\rm d}^3q}{(2\pi)^3}\;
\widetilde{v}(q)\;(q^2)^{\alpha}\widetilde{\chi}(p-q)
\label{oneloopintegral}
\end{equation}
at zero momentum. Recall that the propagators are given by
\Ref{propagators}. The exponential regulator gives a convergent
integral which is evaluated in \Ref{Oneloopintegral}. Second we
have a one loop contribution
\begin{equation}
F^{(m-1,0)}_{(1,\alpha),(m,0)}=
4m(2m-1)\;\widetilde{K}_{1,\alpha}(0)
\label{structurethree}
\end{equation}
as well as a zero loop contribution
\begin{equation}
F^{(m,0)}_{(1,\alpha),(m,0)}=
4m\;\delta_{\alpha,0}.
\label{structurefour}
\end{equation}
This last pairing also contributes to momentum dependent
higher vertices which we neglect. Two higher vertices
yield upon pairing both a quadratic vertex and higher vertices.
One quadratic term is
\begin{equation}
F^{(1,\gamma)}_{(n,0),(n,0)}=
2n(2n-1)(2n)!\;
\widetilde{K}^{(\gamma)}_{2n-2}(0)
\label{structurefive}
\end{equation}
with the $(2n-2)$--loop (the number of soft propagators $v$)
integral
\begin{equation}
\widetilde{K}_{2n-2}(p)=
\widetilde{v}\star\cdots\star\widetilde{v}\star
\widetilde{\chi}(p)
\label{loopintegral}
\end{equation}
expanded into
\begin{equation}
\widetilde{K}_{2n-2}(p)=
\sum_{\alpha=0}^{\infty} (p^2)^{\alpha}
\widetilde{K}^{\alpha}_{2n-2}(0)
\end{equation}
at zero momentum. A second quadratic term is
\begin{equation}
F^{(1,\gamma)}_{(n,0),(n-1,0)}=
(n-1)(2n)!\;\widetilde{K}_{2n-2}(0)\;\delta_{\gamma,0}.
\label{structuresix}
\end{equation}
This second term is exactly local. The general higher
vertex content of the pairing of two higher vertices is
summarized in
\begin{eqnarray}
F^{(l,0)}_{(n,0),(m,0)}&=&
\frac{(2n)!\;(2m)!}{(n+m-l-1)!(m+l-n)!(l+n-m)!}\times
\nonumber\\& &\quad
\widetilde{K}_{n+m-l-1}(0)\;
\Theta_{n+m,l+1}\;\Theta_{m+l,n}\;\Theta_{l+n,m}.
\label{structureseven}
\end{eqnarray}
This last set of structure constants \Ref{structureseven}
alone defines a local approximation for the renormalization
group fixed point. As mentioned above, the general outcome
of the pairing of two higher vertices also contains momentum
dependent vertices which are not encorporated in
\Ref{structureseven}. All other structure constants
between vertices in $\Ind$ are zero. 

The one loop integral \Ref{oneloopintegral} is evaluated to
\begin{equation}
\widetilde{K}_{1,\alpha}(0)=
\frac{1}{(8\pi)^2}\;2^{1/2-\alpha}\;
\Gamma (\alpha+1/2).
\label{Oneloopintegral}
\end{equation}
The $l$--loop integral \Ref{loopintegral} is computed as
a function of the external momentum squared to
\begin{equation}
\widetilde{K}_l(p)=
(4\pi)^{-3l/2}\int_{1}^{\infty}{\rm d}\alpha_1
\cdots{\rm d}\alpha_l\;
A^{-3/2}\;\exp\left(\frac{-B}{A}p^2\right)
\end{equation}
with the abbreviations
\begin{equation}
A=\sum_{m=1}^{l+1}\prod_{n\neq m}\alpha_n,\quad
B=\prod_{m=1}^{l}\alpha_m,
\end{equation}
where $\alpha_{l+1}=1$. Its momentum derivatives at zero
can be reduced further to a one dimensional integral
\begin{equation}
\widetilde{K}_{l}^{(\alpha)}(0)=
\frac{1}{(8\pi)^l}\;
\frac{(-1)^\alpha}{\alpha !\Gamma (\alpha+3/2)}\;
\int_{0}^{\infty}{\rm d}x\;
x^{\alpha+1/2}\;e^{-x}
\left\{\frac{{\rm erf}(\sqrt{x})}{\sqrt{x}}\right\}^{l}.
\end{equation}
This remaining integral can be done explicitely at least
in the one--loop case. We evaluated it in the general case
numerically to high accuracy (45 digits) on the computer.
\section{Eigenvalue problem for critical indices}
\label{ev1}
The fixed point equation \Ref{fixed} comes together with an
eigenvalue problem
\begin{equation}
\left[\dil -\lambda\right]\; W(\phi)=
2\;\bra{V(\phi)}{W(\phi)},
\label{eigenvalueproblem}
\end{equation}
defining scaling fields $W(\phi)$ and their anomalous dimensions
$\lambda$. We emphasize that $W(\phi)$ is a composite field of $\phi$.
The spectrum of anomalous dimensions is the object of principle
interest associated with a fixed point. It directly determines
the critical exponents, see Wilson and Kogut \cite{WK74}.

The interpolation \Ref{better} is accompanied by an interpolation
of \Ref{eigenvalueproblem}, given by
\begin{equation}
\left[\dil -\lambda (z)\right]\; W(\phi,z)=
2\;z\;\bra{V(\phi,z)}{W(\phi,z)}.
\label{intereigen}
\end{equation}
Eq. \Ref{intereigen} can be solved by means of perturbation theory.
We expand not only the interaction \Ref{pertur}, but also the
scaling field and its anomalous dimension into power series in the
interpolation parameter,
\begin{equation}
W(\phi,z)=\sum_{r=0}^{\infty}\; z^r\;\eig{r},\quad
\lambda(z)=\sum_{r=0}^{\infty}\; z^r\;\spe{r}.
\label{eigenseries}
\end{equation}
We interpret \Ref{eigenseries} in the
sense of a formal power series. It yields a solution to
\Ref{intereigen} if the coefficients satisfy the system of
differential equations
\begin{equation}
\dil\eig{r}-\sum_{s=0}^{r}\spe{s}\eig{r-s}=
2\sum_{s=0}^{r-1}\bra{\fix{s}}{\eig{r-1-s}}.
\label{spectrumseries}
\end{equation}
This system can be integrated iteratively. To order zero,
\Ref{spectrumseries} becomes the eigenvalue problem
\begin{equation}
\left[\dil -\spe{0}\right]\; \eig{0}=0.
\end{equation}
The zeroth order $\eig{0}$ thus has to be a scaling field of the
trivial fixed point, and $\spe{0}$ has to be its scaling dimension.
With each scaling field of the trivial fixed point is therefore
associated in perturbation theory a scaling field of the non--trivial
fixed point.

Let us consider for definiteness the perturbation associated with
a mass term
\begin{equation}
\eig{0}=\obs{1}{0},\quad\obs{1}{0}=\int{\rm d}^3x\;\phi(x)^2.
\label{massterm}
\end{equation}
Then the zeroth order eigenvalue is of course $\spe{0}=\sigma_{1,0}=2$. 
As a perturbation of the non--trivial fixed point, \Ref{massterm} turns 
out to be relevant. The associated non--trivial renormalized trajectory 
in the sense of \cite{W96} describes the renormalization group flow of a 
non--trivial massive field theory. Associated with it is the critical 
exponent
\begin{equation}
\nu =\frac{1}{\lambda}.
\label{nu}
\end{equation}
The mass perturbation \Ref{massterm} is non--degenerate in the sense 
that the kernel of $\dil -2$ is one dimensional. The formal orthogonal 
projector on this one dimensional kernel is $\pro{1}{0}$. Another
non--degenerate perturbation is the scaling field associated with
$\obs{2}{0}$. The ones associated with $\obs{1}{1}$ and $\obs{3}{0}$
on the other hand form a degenerate marginal duplet. We will
restrict our attention to the non--degenerate case for the sake
of notational economy. The kernel of $\dil -\spe{0}$ will thus be
assumed to be one dimensional. The formal orthogonal projector on
this rank one kernel will be denoted by $\Ker$. To first order
\Ref{spectrumseries} becomes the differential equation
\begin{equation}
\left[\dil -\spe{0}\right]\;\eig{1}-
\spe{1}\eig{0}=2\bra{\fix{0}}{\eig{0}}.
\label{firstspectrum}
\end{equation}
The first order correction to the eigenvalue follows from
\Ref{firstspectrum} by projection with $\Ker$. We have
that
\begin{equation}
\spe{1}\;\eig{0}=
-2\Ker\bra{\fix{0}}{\eig{0}}.
\label{sigmaone}
\end{equation}
Spelled out explicitely for the mass perturbation \Ref{massterm},
eq. \Ref{sigmaone} says that
\begin{equation}
\spe{1}=-48\;\widetilde{\chi}\star\widetilde{v}(0)\;
\cou{2}{0}{0}=\frac{-1}{3}.
\label{firstnu}
\end{equation}
It is amusing that this first order correction can be inferred
without having to compute the convolution integral, because the 
convolution integral in \Ref{firstnu} is canceled exactly by the one in
\Ref{nontrivial}. Next we impose the normalization condition
that
\begin{equation}
\Ker\eig{1}=0.
\label{spectrumnormalization}
\end{equation}
This condition is appropriate in the non--degenerate case because
\Ref{sigmaone} already takes care of the $\Ker$--information
contained in \Ref{firstspectrum}. The orthogonal complement is
then integrated as in the case of the interaction. The outcome is
\begin{equation}
\Ker^{\perp}\eig{1}=
2\left[\dil -\spe{0}\right]^{-1}
\Ker^{\perp}\bra{\fix{0}}{\eig{0}}.
\end{equation}
This scheme carries on immediately to every order of interpolation
expansion. Projecting \Ref{spectrumseries} to $\Ker$, we first
deduce that
\begin{equation}
\spe{r}\eig{0}=
2\sum_{s=0}^{r-1}\Ker\bra{\fix{s}}{\eig{r-1-s}}.
\end{equation}
This equation determines the order $r$ eigenvalue in terms of lower
order data. Generalizing \Ref{spectrumnormalization}, we impose the
normalization condition
\begin{equation}
\Ker\eig{r}=0
\end{equation}
for $r\geq 1$. Then to order $r$ we are left with the computation
of
\begin{eqnarray}
\Ker^{\perp}\eig{r}&=&
\left[\dil -\spe{0}\right]^{-1}
\Biggl\{\sum_{s=1}^{r-1}\spe{s}\Ker^{\perp}\eig{r-1}+
\nonumber\\& &\quad
2\sum_{s=0}^{r-1}\Ker^{\perp}\bra{\fix{s}}{\eig{r-1-s}}.
\Biggr\}
\label{spectrumintegral}
\end{eqnarray}
This scheme iterates to every order of interpolation expansion.
Recall that the inverse of the dilatation generator in
\Ref{spectrumintegral} involves the integration of a first order
partial differential equation. As in the case of the fixed point,
the explicit computation of this program to very high orders
requires considerable computational resources. In this paper we
restrict our attention to a partial resummation by means of our
coordinate representation.
\section{Eigenvalue problem in coordinates}
\label{ev2}
In this section we perform the interpolation expansion for
the eigenvalue problem \Ref{intereigen} in the coordinate
representation. The coordinate representation for the scaling
fields reads
\begin{equation}
W(\phi,z)=\sum_{i\in\Ind}\Obs{i}\;W^{i}(z).
\label{cooeig}
\end{equation}
In the coordinate representation, the eigenvalue problem
\Ref{intereigen} becomes a set of algebraic equations
\begin{equation}
\left(\sigma_{k}-\lambda (z)\right)
W^{k}(z)=2z\sum_{i,j\in\Ind}F^{k}_{i,j}\;V^{i}(z)\;
W^{j}(z).
\label{eigenalgebra}
\end{equation}
It can be solved recursively in an interpolation expansion
\begin{equation}
W^{k}(z)=\sum_{r=0}^{\infty}\;z^r\;\Coo{k}{r}.
\label{scalepert}
\end{equation}
The power series \Ref{coeffs}, \Ref{eigenseries}, and
\Ref{scalepert} yield a solution to \Ref{eigenalgebra} provided
that the coefficients satisfy
\begin{equation}
\left(\sigma_k-\spe{0}\right)\;\Coo{k}{r}-
\spe{r}\;\Coo{k}{0}=
\sum_{s=1}^{r-1}\spe{s}\;\Coo{k}{r-s}+
2\sum_{s=0}^{r-1}\sum_{i,j\in\Ind}
F^{k}_{i,j}\;\Cou{i}{s}\;\Coo{j}{r-1-s}.
\label{scaleeigen}
\end{equation}
As in the case of the interaction, the system of equations
\Ref{scaleeigen} can be organized into a recursion relation. To
order zero \Ref{scaleeigen} reads
\begin{equation}
\left(\sigma_k-\spe{0}\right)\Coo{k}{0}=0.
\end{equation}
It tells us that we should select one of the $k\in\Ind$ as
zeroth order eigenvector. We choose $\underline{1}=(1,0)$
for definiteness. Then the zeroth order is
\begin{equation}
\Coo{k}{0}=\delta_{\underline{1},k},\quad
\spe{0}=\sigma_{\underline{1}}=2.
\end{equation}
The only $k$ with $\sigma_k=2$ is $k=\underline{1}$. We will
again restrict our attention to this non--degenerate case. The
below recursion relation is valid for general non--degenerate
perturbations, with minor notational changes. The first order
equation in the system \Ref{scaleeigen} is given by
\begin{equation}
\left(\sigma_k-\spe{0}\right)\;\Coo{k}{1}-
\spe{1}\Coo{k}{0}=2\;F^{k}_{\underline{2},\underline{1}}
\Cou{\underline{2}}{0}.
\end{equation}
Therefrom it follows that the first order correction to the
eigenvalue is in the coordinate representation
\begin{equation}
\spe{1}=-2\;F^{\underline{1}}_{\underline{2},\underline{1}}
\Cou{\underline{2}}{0}.
\end{equation}
We remark that in the degenerate case, the degeneracy is
typically lifted by the first order correction to the
eigenvalue. The other coefficients to first order are
\begin{equation}
\Coo{\underline{1}}{1}=0
\end{equation}
and, for $k\in\Ind\setminus\{\underline{1}\}$,
\begin{equation}
\Coo{k}{1}=
\frac{2\;F^{k}_{\underline{2},\underline{1}}}{\sigma_k-\spe{0}}.
\end{equation}
This computation generalizes immediately to higher orders. The
formula for the order $r$ eigenvalue in terms of lower order data
is
\begin{equation}
\spe{r}=-2\sum_{s=0}^{r-1}\sum_{i,j\in\Ind}
F^{\underline{1}}_{i,j}\;\Cou{i}{s}\;\Coo{j}{r-1-s}.
\label{recursion1}
\end{equation}
The order $r$ eigenvector is then given by
\begin{equation}
\Coo{\underline{1}}{r}=0,
\label{recursion2}
\end{equation}
for $r\geq 1$, together with
\begin{equation}
\Coo{k}{r}=\frac{1}{\sigma_{k}-\spe{0}}
\Biggl\{\sum_{s=1}^{r-1}\;\spe{s}\;\Coo{k}{r-s}+
2\sum_{s=0}^{r-1}\sum_{i,j\in\Ind}F^{k}_{i,j}\;
\Cou{i}{s}\;\Coo{j}{r-1-s}\Biggr\},
\label{recursion3}
\end{equation}
for $k\in\Ind\setminus\{\underline{1}\}$. Eq. \Ref{recursion1},
\Ref{recursion2}, and \Ref{recursion3} define a recursive
perturbation expansion for the critical indices of the non--trivial
fixed point.
\section{Computation of the recursions}

We computed the $z$--expansion for the potential recursively by 
means of (\ref{couprecursion}) and (\ref{otherrecursion}), and 
for the eigenvalue problem by means of (\ref{recursion1}) and 
(\ref{recursion3}) using computer algebra. We restricted our 
attention to the case of three dimensions. It turned out to be
crucial to compute the structure coefficients to high accuracy.
We calculated them to an accuracy of 45 digits with Maple V.
The perturbation expansion was performed up to a maximal order
of 25. The derivative expansion was performed up to $\am=20$
orders of $p^2$ in the 2-point vertex. Table \ref{table1} shows
the series for the $\phi^4$--coupling both in the ultra--local
approximation $\am=0$ and for $\am=4$ up to the order $z^{11}$.
\begin{table}[htpb]
\begin{center}
\leavevmode
 \begin{tabular}[r]{|r|r@{$\cdot$}l|r@{$\cdot$}l|}\hline
\mc{1}{|c|}{$n$} &
\mc{2}{c|}{$\cou{2}{0}{n}$, $\am=0$} &
\mc{2}{c|}{$\cou{2}{0}{n}$, $\am=4$} \\ \hline
   0 & 2.1874 & $10^{-1}$ & 2.1874 & $10^{-1}$ \\
   1 & 4.5814 & $10^{-1 }$& 4.5814 & $10^{-1}$ \\
   2 & -8.7171 & $10^{-1 }$& -8.6761 & $10^{-1}$ \\
   3 & 4.6575 & $10^{0 }$& 4.6815 & $10^{0}$ \\
   4 & -4.0553 & $10^{1 }$& -4.0546 & $10^{1}$ \\
   5 & 4.2980 & $10^{2 }$& 4.2992 & $10^{2}$ \\
   6 & -5.2117 & $10^{3 }$& -5.2130 & $10^{3}$ \\
   7 & 7.0118 & $10^{4 }$& 7.0133 & $10^{4}$ \\
   8 & -1.0267 & $10^{6 }$& -1.0269 & $10^{6}$ \\
   9 & 1.6155 & $10^{7 }$& 1.6158 & $10^{7}$ \\
   10 & -2.7080 & $10^{8 }$& -2.7084 & $10^{8}$ \\
   11 & 4.8059 & $10^{9 }$& 4.8066 & $10^{9}$ \\ \hline
 \end{tabular}
 \parbox[t]{\textwidth}
  {
   \caption[]{\label{table1}
   \sl Examples for the behaviour of the expansion coefficients.}
  }
\end{center}
\end{table}
The coefficients prove to increase in absolute value proportional
to $C^n n!$ with some constant $C$. Their signs alternate. From 
this behavior we conclude that the series does not converge but
is Borel summable. A proof of local Borel summability will be 
presented elsewhere. The constant $C$ is related to an instanton
singularity of the Borel transform on the negative real axis. It
can be seen as an accumulation point of poles when the series is
converted into various Pade approximants. The derivative expansion
on the other hand proves to converge. This is illustrated in
table \ref{table2} for two values of $\am$.
\begin{table}[htpb]
\begin{center}
\leavevmode
\begin{tabular}[r]{|r|r@{$\cdot$}l|r@{$\cdot$}l|}\hline
\mc{1}{|c|}{$\alpha$} &
\mc{2}{c|}{$\cou{2}{\alpha}{10}$, $\am=5$} &
\mc{2}{c|}{$\cou{2}{\alpha}{10}$, $\am=10$} \\ \hline
0 & -1.08730107 & $10^{5}$ & -1.08730280 & $10^{5 }$\\
1 & -7.30254527 & $10^{4 }$& -7.30254464 & $10^{4 }$\\
2 & 9.42673139 & $10^{3 }$& 9.42673053 & $10^{3 }$\\
3 & -1.47739400 & $10^{3 }$& -1.47739385 & $10^{3 }$\\
4 & 2.14057814 & $10^{2 }$& 2.14057791 & $10^{2 }$\\
5 & -2.79810089 & $10^{1 }$& -2.79810056 & $10^{1 }$\\
6 & 0.00000000 & $10^{0 }$& 3.33201984 & $10^{0 }$\\
7 & 0.00000000 & $10^{0 }$& -3.70314164 & $10^{-1 }$\\
8 & 0.00000000 & $10^{0 }$& 3.99772841 & $10^{-2 }$\\
9 & 0.00000000 & $10^{0 }$& -4.43094961 & $10^{-3 }$\\
10 & 0.00000000 & $10^{0 }$& 5.30798357 & $10^{-4 }$\\ \hline
 \end{tabular}
 \parbox[t]{\textwidth}
  {
   \caption[]{\label{table2} \sl
    Examples for the behaviour of the mass coefficients at order
    $z^{10}$ for $\am=5$ and $\am=10$.}
  }
\end{center}
\end{table}
We note in passing that the difference between $\am=5$ and $\am=10$ 
is small.

The spectrum of the non--trivial fixed is computed along the strategy
explained in section \ref{ev1} and \ref{ev2}. It requires as an input
the fixed point interaction in $z$--expansion. We evaluated it for 
all values of $\am$ inbetween zero and twenty. In the following we 
will concentrate on an estimate of the critical index $\nu$ (\ref{nu})
by resummation of the series for all these twenty one approximations.
We computed the series by means of (\ref{recursion1}) and 
(\ref{recursion3}) to order twenty five of $z$--expansion. 

\begin{figure}[htbp]
  \begin{center}
    \leavevmode
    \epsfxsize=12cm
    \epsffile{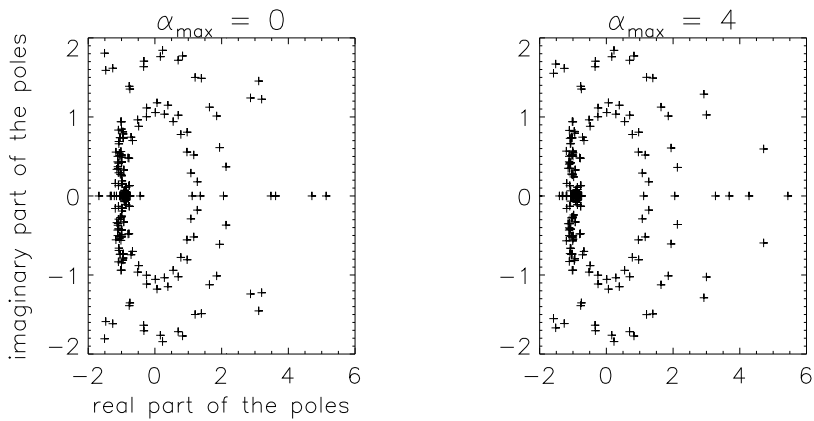}
    \caption{\sl Radius of convergence of $B\lambda$ by the \pade method for
      $\am=0$ and $\am=4$.}
    \label{fig1}
  \end{center}
\end{figure}

Table \ref{table3} shows as an example the series for the eigenvalue 
$\lambda$ in the ultra--local case $\am=0$ and in the case of $\am=4$ 
up to the order twelve of perturbation theory.
\begin{table}[htpb]
\begin{center}
\leavevmode
\begin{tabular}[r]{|r|r@{$\cdot$}l|r@{$\cdot$}l|}\hline
\mc{1}{|c|}{$n$} &
\mc{2}{c|}{$\la{n}$, $\am=0$} &
\mc{2}{c|}{$\la{n}$, $\am=4$} \\ \hline
0 & 2.000000 &$10^{0}$  & 2.000000 &$10^{0}$  \\
1 & -3.333333 &$10^{-1}$  & -3.333333 &$10^{-1}$  \\
2 & -3.490659 &$10^{-1}$  & -3.490659 &$10^{-1}$  \\
3 & 1.148993 &$10^{0}$  & 1.159189 &$10^{0}$  \\
4 & -7.414413 &$10^{0}$  & -7.369227 &$10^{0}$  \\
5 & 6.358855 &$10^{1}$  & 6.358630 &$10^{1}$  \\
6 & -6.649232 &$10^{2}$  & -6.646081 &$10^{2}$  \\
7 & 7.999490 &$10^{3}$  & 7.996744 &$10^{3}$  \\
8 & -1.070838 &$10^{5}$  & -1.070532 &$10^{5}$  \\
9 & 1.562548 &$10^{6}$  & 1.562166 &$10^{6}$  \\
10 & -2.452524 &$10^{7}$  & -2.452002 &$10^{7}$  \\
11 & 4.103373 &$10^{8}$  & 4.102601 &$10^{8}$  \\
12 & -7.271917 &$10^{9}$  & -7.270698 &$10^{9}$  \\ \hline
 \end{tabular}
 \parbox[t]{\textwidth}
  {
   \caption[]{\label{table3}\sl
    Series coefficients for $\lambda$ up to
    order 12 for $\am=0$ and $\am=4$.}
  }
\end{center}
\end{table}
Again the series alternate, and the coefficients grow in absolute 
value as $C^n n!$. The series are therefore not expected to converge. 
We remark that a proof thereof is however missing. The Borel transform 
of a series with this asymptotics has a finite radius of analyticity
$R_{\am}$. It is determined by an instanton 
singularity on the negative real axes of the complex Borel plane. 
This radius of analyticity is an interesting quantity. It can be 
investigated by a number of methods, see \cite{DI89} and references
therein. One of them is the
\pade method. Recall that the \pade approximant of order $(l,m)$ for
a function $f$ is a rational function $f_{l,m}(z)=\frac{P_l(z)}{Q_m(z)}$.
Here $P_l$ and $Q_m$ are polynomials of degree $l$ and $m$ respectively,
determined such that the taylor expansions of $f$ and $f_{l,m}$ agree up 
to order $z^{l+m}$. One then observes that the poles of the various
possible \pade approximants accumulate around a cut or a singularity of 
$f$. With the \pade method 
we found
\begin{equation}
  \label{conv}
  R_{\am} = 0.88 \pm 0.02 \, ,
\end{equation}
with no significant dependence of $\am$. Figure \ref{fig1} shows a
plot of all poles of all \pade approximants $(B\lambda)_{l,m}$ with
$l+m=25$ in the complex Borel plane for the two cases $\am=0$ and 
$\am=4$ respectively.
Here $B\lambda$ denotes the Borel transform of $\lambda$. As expected, the 
poles accumulate on the negative real axes. Notice however that there are 
many spurious singularities on and nearby the positive real axes. These
spurious poles endanger the inverse Borel transform as a contour integral
along the positive real axis. The pictures for $\am=0$ and $\am=4$ show 
tiny differences. For instance, the poles on the positive real axis are
not on fixed locations and can therefore be regarded as spurious.
\section{Determination of $\nu$}
\label{sec:nu}
  
To compute the value of the critical index $\nu$, we have to evaluate 
the $z$--expansion at $z=1$. Naive evaluation does not give a meaningful
answer since the expansion does not converge. Therefore we had to rely
on resummation technology. A review of series resummation and references
to the original literature is given in \cite{ZJ89} and \cite{DI89}. 
We tried four standard 
methods and compared the results.

First we computed (for all values of $\am$ between zero and twenty) all 
\pade approximants $(\lambda)_{l,m}$ with $l+m \leq 25$, and evaluated 
them at $z=1$. These values are conveniently displayed in a \pade table 
($l$-$m$ grid). To get an idea for the value of $\lambda$ and 
an estimate for the error we computed the mean value and deviation 
for the lines of fixed order in $z$ ($l+m=\mbox{const}$) in these
diagrams after having discarded all values below a
lower value $\lambda_{min}$ and above an upper value $\lambda_{max}$.
The idea thereof is that large 
deviations come from spurious singularities. We were careful not to choose 
the window too narrow. Our error estimate should be regarded as rather
pessimistic.
If these mean values converge with increasing order
in $z$ we use them and
an inspection of the whole table to find an estimate
for the value of $\nu$.


In the second method (Dlog) one computes the \pade approximants
for the logarithmic derivative $\frac{\lambda'(z)}{\lambda(z)}$.
$\lambda$ is then reconstructed as the exponential of an
integral
\begin{equation}
  \label{dlog}
  \lambda_{l,m} = \lambda(0) e^{\int_0^1 dz
    \left( \frac{d}{dz} \log\lambda(z)\right)_{l,m}} \, .
\end{equation}
The integration can be performed numerically to high accuracy.
The Dlog method is particularly efficient when the singularity is of 
the type $\lambda(z) = \frac{A}{(x-x_c)^{\gamma}}$ with a nonintegral
exponent $\gamma$.

The third proposal is to use a \pade approximants for the Borel 
transform of the series. The Borel transform of a
power series $f(x) = \sum_{n\geq 0} f_n x^n$ is defined by
\begin{equation}
  \label{bp}
  (B f)(z) = \sum_{n\geq 0} \frac{f_n}{n!} z^n \, .
\end{equation}
The Borel transform of power series with 
finite radius of convergence defines an analytic continuation of
the function to a maximal simplex through the integral
\begin{equation}
  \label{bpback}
  f(x) = \int_0^{\infty} dt e^{-t} (B f)(x t).
\end{equation}
Again we get \pade tables of approximants for $\lambda$ by numerically
integrating this back transformation for various \pade approximants of 
$B\lambda$.

The off diagonal estimates in these tables can be improved by using
information on the analyticity properties of the Borel transform. 
$B f$ could for instance have a cut along $(-\infty, -R]$ on the negative 
real axes. Let us assume that this is indeed the case (with $R=0.88\pm 
0.02$). Then the cut plane can be mapped conformally via
\begin{equation}
  \label{conf}
  u(z) = \frac{\sqrt{z/R+1}-1}{\sqrt{z/R+1}+1} \, ;
  \quad
  z = \frac{4Ru}{(1-u)^2}
\end{equation}
onto the unit circle. Under this mapping $(Bf)(z)$ transforms to 
$(\tilde{B f})(u)$. We then use \pade approximants for the mapped series.
A \pade table for $\lambda$ is obtained via the inverse transformation
\begin{equation}
  \label{bpconfback}
  \lambda_{l,m} = 4R\int_0^1 du \frac{1+u}{(1-u)^3}
  e^{-\frac{4Ru}{(1-u)^2}}(\tilde{B f})_{l,m}(u) \, .
\end{equation}
The outcome of this method relies on a careful estimate of the radius of 
convergence of the Borel transform. For each $\am$, we calculated three 
estimates for $\lambda$, one for our estimated value of $R$ and one for 
$R+\Delta R$ and $R-\Delta R$ respectively, where $\Delta R$ means the
error in our estimate for the error of the radius. The inspection of all 
three \pade tables yields $\lambda$ and an error estimate.

We also tried out inhomogenous differential approximants, but we 
could see no improvement as compared with \pade or Dlog \pade
approximants. The integration of the differential equations in this 
method turned out to be both time consuming and fragile due to 
the poles close to the origin. 

\begin{figure}[htbp]
  \begin{center}
    \leavevmode
    \epsfxsize=14cm
    \epsffile{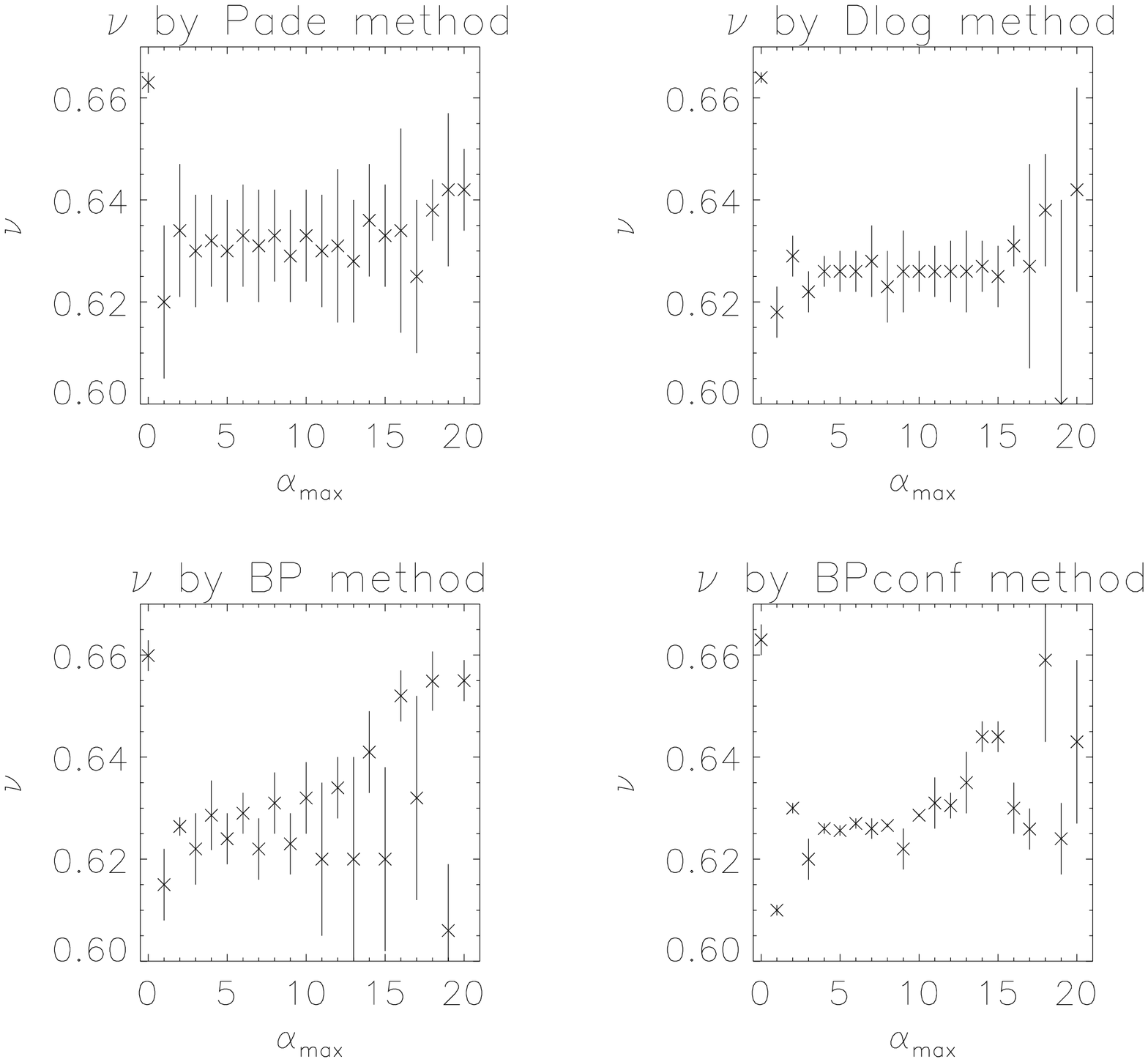}
    \caption{\sl The critical index $\nu$ as a function of
      the order of the derivative expansion of the
      2-point function $\am$ for the four extrapolation
      methods.}
    \label{fig2}
  \end{center}
\end{figure}

In table \ref{table4} we summarize our results for $\nu$ for the different 
values of $\am$. The errors refer as usually to the last digit.
\begin{table}[htpb]
\begin{center}
\leavevmode
\begin{tabular}[r]{|r|l|l|l|l|}\hline
\mc{1}{|c|}{$\am$} &
\mc{1}{c|}{$\nu$ \pade} &
\mc{1}{c|}{$\nu$ Dlog} &
\mc{1}{c|}{$\nu$ BP} &
\mc{1}{c|}{$\nu$ BPconf} \\ \hline
  0 & 0.6630(20) & 0.6640(10) & 0.6599(30) & 0.6630(30) \\
  1 & 0.6200(150) & 0.6180(50) & 0.6150(70) & 0.6100(10) \\
  2 & 0.6340(130) & 0.6290(40) & 0.6264(18) & 0.6300(10) \\
  3 & 0.6300(110) & 0.6220(40) & 0.6220(70) & 0.6200(40) \\
  4 & 0.6320(90) & 0.6260(30) & 0.6286(68) & 0.6260(10) \\
  5 & 0.6300(100) & 0.6260(40) & 0.6240(50) & 0.6256(10) \\
  6 & 0.6330(100) & 0.6260(40) & 0.6290(40) & 0.6270(10) \\
  7 & 0.6310(110) & 0.6280(70) & 0.6220(60) & 0.6260(20) \\
  8 & 0.6330(90) & 0.6230(70) & 0.6310(60) & 0.6266(4) \\
  9 & 0.6290(90) & 0.6260(80) & 0.6230(60) & 0.6220(40) \\
  10 & 0.6330(90) & 0.6260(40) & 0.6320(70) & 0.6286(3) \\
  11 & 0.6300(110) & 0.6260(50) & 0.6200(150) & 0.6310(50) \\
  12 & 0.6310(150) & 0.6260(60) & 0.6340(60) & 0.6305(25) \\
  13 & 0.6280(120) & 0.6260(80) & 0.6200(200) & 0.6350(60) \\
  14 & 0.6360(110) & 0.6270(50) & 0.6410(80) & 0.6440(30) \\
  15 & 0.6330(100) & 0.6250(60) & 0.6200(180) & 0.6440(30) \\
  16 & 0.6340(200) & 0.6310(40) & 0.6520(50) & 0.6300(50) \\
  17 & 0.6250(150) & 0.6270(200) & 0.6320(200) & 0.6259(40) \\
  18 & 0.6380(60) & 0.6380(110) & 0.6549(58) & 0.6590(160) \\
  19 & 0.6420(150) & 0.6000(400) & 0.6060(130) & 0.6240(70) \\
  20 & 0.6420(80) & 0.6420(200) & 0.6550(40) & 0.6430(160) \\ \hline
 \end{tabular}
 \parbox[t]{\textwidth}
  {
   \caption[]{\label{table4}\sl
     Results for the critical exponent $\nu$ with the
     \pade method, the Dlog \pade method (Dlog), the Borel \pade method (BP)
     and the Borel \pade method with conformal mapping (BPconf) for
     various orders of the derivative expansion of the 2-point vertex.}
  }
\end{center}
\end{table}
We come to the following conclusions. The \pade method is the least precise 
one with an error of about 0.01. From it we can get an idea about the value
of $\nu$, but no accurate estimate. The errors of the Dlog \pade method 
(Dlog) and the Borel \pade method (BP) are of comparable size. The Borel 
\pade method with conformal mapping (BPconf) has the least errorbars. At higher 
orders of the derivative expansion of the 2-point vertex the errors increase
significantly. To display this effect, we have plotted the data of table 
\ref{table4} in figure \ref{fig2}.
One can see that the values for $\nu$ oscillate around a mean value. 
Up to a certain order this sequence seems to converge. Thereafter, the 
difference between $\nu(\am)$ and $\nu(\am + 1)$ and the error grows. 

We believe this effect to be the consequence of a numerical instability.
In high orders of derivative expansion and high orders of perturbation
theory one is dealing with numbers of enormously varying magnitudes
(in our case one hundred orders). In practice we computed our series
to an accuracy of 45 digits, and a problem arises in the cancellation
of large numbers in the course of the recursion. A more destructive
explanation would be that the resummation fails to produce a convergent
derivative expansion, or even more desastrous that the non--perturbative
kernels are not analytic functions of the momenta. The final answer to
this question can only be given on the basis of a non--perturbative 
construction of the fixed point and is outside the scope of this paper.
Our insight comes from the evaluation of various approximants to 
different orders of accuracy.

We confine our further discussion to those values of $\am$ which lie 
before the onset of instability. The Dlog method and the BPconf method 
both yield nearly constant values for $\nu$ at orders between $\am=4$
and $\am=12$ and between $\am=4$ and $\am=8$ respectively. We propose 
this value to be the limit of $\nu$ at arbitrary order of the derivative 
expansion.

Consider the data for the ultra--local case $\am=0$ and to first order
$\am=1$ of derivative expansion. For the ultra--local case,
which can be compared with the hierarchical model ($\nu=0.6501625$,
\cite{KW88}) we find 
$\nu=0.6625(33)$ which is bigger than the full critical index.
Disregarding the pure \pade estimate, we get for $\am=1$
the result $\nu = 0.6144(62)$.  This value is considerably lower than 
the value at $\am=0$ and even lower than the full critical index.
In view of the tiny differences between the fixed point coefficients
at $\am=0$ and $\am \neq 0$, we find this surprising. Compare for
example the coefficients in table \ref{table1}.  

At higher orders of derivative expansion, the values for $\nu$ oscillate 
and converge to a mean value. The limit value has been determined
as the mean values of $\nu$ over the nearly constant plateaus.
As best estimate for the BPconf method we get 
$\nu = 0.6262(13)$. The Dlog method yields $\nu=0.6259(57)$. These results 
should be compared with the critical index $\nu$ of the three dimensional 
Ising model in the literature. In table \ref{table5} we list a few
results for $\nu$.
\begin{table}[htbp]
  \begin{center}
    \leavevmode
    \begin{tabular}{|l|l|c|}\hline
      \mc{1}{|c|}{$\nu$} &
      \mc{1}{c|}{Method} &
      \mc{1}{c|}{Literature} \\ \hline
      0.6300(15) & three dimensional renormalization group & \cite{GZJ80}\\
      0.6298(7)  & & \cite{BB85}\\
      0.630      & & \cite{N91}\\ \hline
      0.6305(25) & renormalization group, $\epsilon$-expansion & \cite{GZJ85}\\ \hline
      0.6301     &  high temperature series & \cite{R95}\\
      0.6300(15) & high temperature series for bcc-grid & \cite{NR90}\\ \hline
      0.6289(8)  & Monte-Carlo methods & \cite{FL91}\\
      0.6301(8)  & & \cite{BLH95}\\ \hline
      0.625(1)   & Monte-Carlo renormalization group & \cite{GT96}\\ \hline
      0.626(9)   & Scaling-field method & \cite{NR84}\\ \hline
    \end{tabular}
    \caption{\sl Results for the critical exponent $\nu$ of the full model.}
    \label{table5}
  \end{center}
\end{table}
A comprehensive article on this issue is \cite{BLH95}. It also contains 
an overview of experimental data. With series expansion and Monte-Carlo 
methods one gets $\nu=0.630$. On the other hand the Monte-Carlo 
renormalization group suggests $\nu = 0.625$. This gap is object of 
current discussions. Our value is closest to the value of
\cite{NR84} and \cite{GT96}.

\section{Summary and discussion}
\label{sec:disc}

In this article we investigated a form of Wilsons infinitesimal 
renormalization group. The starting point was equation \Ref{fixed}.
We found a practical way to solve the equation in a systematic manner. 
The central idea was to introduce an interpolating parameter $z$,
which continuously turns on the non--linear term in \Ref{fixed}. 
Everything was expanded in this parameter. The interpolation was
arranged such that the zeroth order is a $\phi^4$--vertex.
The expansion was presented both in a coordinate free representation
and in coordinate form, where the interaction is expanded in a basis of 
vertices. As a basis we advocated the use of a full two point interaction
in derivative expansion together with local vertices of any power of
fields. Derivative interactions of higher powers were neglected. The
basis of interactions came encoded in a system of scaling dimensions and
structure constants. Their evaluation was reduced to a one dimensional
Feynman integral which we evaluated numerically. We reformulated our
expansion into recursive equations for the fixed point interaction,
its scaling fields, and their anomalous dimensions. We performed a
detailed analysis of the series for the critical exponent associated
with a massive perturbation of the fixed point. The result is a new 
and independent calculation of the critical index $\nu$ of the
the three dimensional Ising model. We solved the recursion relations 
for the eigenvalue problem up to high orders and analyzed the resulting 
series by means of four different extrapolation methods. Our best 
estimator for the critical index $\nu$ is $\nu = 0.6262(13)$. We 
compared our results with values for the critical exponent $\nu$ known 
in the literature.

The results encourage us to further investigations. On the menu of
open problems we have the inclusion of momentum dependent higher 
vertices for the scalar model, theoretical estimates on the 
$z$--expansion, and the generalization to vector and matrix models. 
We hope to return with accurate data on their critical properties
by means of $z$--expansion in the near future.

\end{document}